\documentstyle[12pt,aasms4]{article}

\begin{document}

\title {THE DYING-WIND AROUND HD 56126, A POST-AGB CARBON STAR}

\author{M. Jura and C. Chen}
\affil{Department of Physics and Astronomy, University of California,
    Los Angeles CA 90095-1562; jura@clotho.astro.ucla.edu; cchen@astro.ucla.edu}
\begin{center}
and
\end{center}
\author{M. W. Werner}
\affil{Jet Propulsion Laboratory, 264-767, 4800 Oak Grove Dr., Pasadena CA 91109; mwerner@sirtfweb.jpl.nasa.gov}

\begin{abstract}
 We have used the Keck I telescope  to resolve  at  mid-infrared wavelengths 
the dust emission from HD 56126 (IRAS 07134+1005), a post-Asymptotic Giant Branch carbon star with a detached dust shell.  The gross morphology of the image can be explained
by a strong wind which started to die about 1500 years ago.  If the
star had an effective temperature, $T_{*}$, near 2600 K when it was losing ${\sim}$3 ${\times}$ 10$^{-5}$ M$_{\odot}$ yr$^{-1}$, then during the past 1500 yr, the average value of
$dT_{*}/dt$ has been +2.2 K yr$^{-1}$.  With such a time variation of the effective temperature,  the 11.7 ${\mu}$m image can be approximately reproduced if the mass loss 
rate varied as $T_{*}^{-8.26}$, as proposed in recent models for dust-driven winds.
Since the mass loss rate appears to be very sensitive to the effective temperature of the star, we speculate that the observed deviations from spherical symmetry of the dust shell can be explained by plausible variations in the surface temperature of
the mass-losing star caused by rotation and/or magnetic fields.

\end{abstract}
\keywords{stars: AGB and post-AGB; circumstellar matter; winds-outflows}

\section{INTRODUCTION}
Mass loss with rates near ${\sim}$ 10$^{-5}$ M$_{\odot}$ yr$^{-1}$ during the Asymptotic Giant Branch (AGB) phase is important both in
the evolution of the mass-losing star and in replenishing the interstellar medium (see, for example, Habing 1996).  After a star evolves off the AGB, its effective temperature rises, its radius shrinks and the wind dies   
 (Vassiliadis \& Wood 1993, Blocker 1995, Schroder, Winters \& 
Sedlmayr 1999).  

Stars slightly beyond the end of their AGB evolution can be identified by having only relatively cold dust since they are no longer losing much mass.  Therefore, they exhibit a large infrared excess but with  dust  temperatures typically less than 200 K (see, for example, Kwok 1993).  The detached dust shell can be extended over the sky and thus provide the opportunity to examine the history of the star's mass loss.

Meixner et al. (1997, 1999) and Dayal et al. (1998) have acquired mid-infrared images and resolved more
than 10 post-AGB stars or ``transition objects".      
 Even in the best-studied cases, the angular diameter of the dust emission is  not much larger than the characteristic $0{\farcs}9$ angular resolution available at the 4m-class telescopes used for these
previous studies.  
With the 10m Keck telescope, it is possible to achieve significantly higher
angular resolution (see, for example, Jura \& Werner 1999, Jura, Chen \& Werner 2000) and thus develop more exacting tests of models for dying winds. 

      Here, we report the result of using Keck to obtain mid-IR images of HD 56126 (m$_{V}$ = 8.27 mag, spectral type F5I), a post-AGB carbon star with a nebula displaying one of the largest angular sizes in the sample studied by Meixner et al. and extended optical reflection nebulosity as well (Ueta, Meixner \& Bobrowsky 2000). This star is also a  ``21 ${\mu}$m" source (see, for example Kwok, Volk \& Hrivnak 1999), a feature attributed to  TiC (von Helden et al. 2000).
In  Section 2 we present our observations while  in Section 3, we describe our estimates for the physical properties of the circumstellar envelope and the central star.  In section 4, we present a model to explain the gross features of the system while  in Section 5, we discuss models to explain deviations from a spherically symmetric wind. In Section 6 we present our conclusions.  
  
\section{OBSERVATIONS}
Our data were obtained on the night 2000 Feb 20 (UT) at the Keck I telescope 
using the Long Wavelength Spectrometer (LWS) which was built by a team 
led by B. Jones and is described on the Keck web page.  The LWS uses a 
128 ${\times}$ 128 SiAs BIB array with
a pixel scale at the Keck telescope of 0{\farcs}08 and a total field of 
view of 10{\farcs}2 ${\times}$ 10{\farcs}2.  We used the ``chop-nod" 
mode of observing, and 4 different filters: 7.5 - 8.2 ${\mu}$m, 9.4-10.2 ${\mu}$m, 11.2 - 12.2 ${\mu}$m and 18.4-18.9 ${\mu}$m.  We used ${\alpha}$ Boo  for flux and point-spread-function calibrations.   The data were reduced
at UCLA using standard LWS routines.   

The image in the 11.7 ${\mu}$m  filter is presented in 
Figure 1; the images at other wavelengths appear similar.     Probably  because of our higher angular resolution, our measured  peak surface brightness of 4.0 Jy arcsec$^{-2}$ is greater than the peak values of 2.3 Jy arcsec$^{-2}$ and 3.0 Jy arcsec$^{-2}$ found by Meixner et al. (1997) and
Dayal et al. (1998), respectively. Given the ${\sim}$10\% uncertainties in our data, our measured 
flux at 11.7 ${\mu}$m of 28 Jy is consistent  with the values for the flux of 30 ${\pm}$ 3 Jy (Dayal et al. 1998), 25.7 ${\pm}$ 2.6 Jy (Meixner et al.  1997) 25 ${\pm}$ 3 Jy (Justtanont et al. 1996) and 25 Jy (IRAS).

Beyond 1{\farcs}5 from the center of the nebula, the isophotes are smooth and elliptically shaped with the major axis lying near Position Angle 45$^{\circ}$ and being about 7\% greater than the minor axis.  A hint of ellipticity at this position angle is also seen in the mid-IR images of Meixner et al. (1997) and
Dayal et al. (1998) as well as in the optical reflection nebulosity presented
by Ueta et al. (2000).
In the inner 1{\farcs}5 there is a marked change in the morphology of the dust with two bright blobs
lying near the minor axis of the ellipse defined by the outer contours.  The bright blob at about 1{\arcsec} to the southeast of the center of the image is seen in the data presented both by Dayal et al. (1998) and Meixner et al. (1997).  The maximum intensity in the bright blob is about a factor of 1.5 greater than the maximum intensity in the faint blob.
Our image at 8${\mu}$m shows that HD 56126 lies within ${\pm}$ 0{\farcs}2 of the center of 
the dust ring defined by the outer, elliptical isophotes.     
    
\section{STELLAR AND CIRCUMSTELLAR PARAMETERS}
The observed pulsation properties of HD 56126 can be used to place the star
in the H-R diagram and to estimate its mass.  
Since HD 56126 is variable with a period of 36.8 days (Barthes et al. 2000), it is plausible that the mass of the star is near 0.6 M$_{\odot}$ and the time-averaged luminosity is near 6600 L$_{\odot}$ (Tuchman et al. 1993, Jeannin et al. 1997, Barthes et al. 2000).  From the observed time-averaged flux (see Meixner et al. 1997), we therefore infer a distance of 2300 pc.  The time-averaged effective temperature, $T_{*}$, and radius,  $R_{*}$, are 5900 K and 5.5 ${\times}$ 10$^{12}$ cm, respectively (Barthes et al. 
2000).  Although carbon, nitrogen and oxygen have nearly solar abundances, since the iron group elements have abundances about 0.1 of solar (Parthasarathy, Garcia Lario \& Pottasch 1992, Klochkova 1995, Van Winckel \& Reyniers 2000), HD 56126 probably is a member of the thick disk and its main sequence progenitor probably had a mass near 1.1 M$_{\odot}$.  Therefore, it is likely that HD 56126 has already experienced a total mass loss of 0.5 M$_{\odot}$.  Evidence that HD 56126 is a post-AGB star
is provided by the greater than solar abundance of the s-process elements
such as Sr, Y, Zr and Ba.  

Below, we propose that ${\sim}$1500 years ago, HD 56126 was losing ${\sim}$3 ${\times}$ 10$^{-5}$ M$_{\odot}$ yr$^{-1}$.  In this phase, its effective temperature was probably near 2600 K (Arndt et al. 1997).  Therefore,
during the past 1500 yr, the average value of $dT_{*}/dt$ has been +2.2 K yr$^{-1}$. The star probably maintained a constant luminosity during its post-AGB evolution (see, for example, Vassiliadis \& Wood 1993) so that as the temperature
increased, the radius decreased from 2.8 ${\times}$ 10$^{13}$ cm to its current
value.     

Although, as inferred  from the profile of the CO (2-1) circumstellar emission line,  most of the gas outflow is at 10.7 km s$^{-1}$ (Knapp et al. 1998), there are significant deviations from this value.  Crawford \& Barlow (2000)
have obtained optical spectra with a resolution of 860,000, and they found three distinct components in the K I circumstellar shell lines with a net spread
in the  velocity of the circumstellar gas of 2 km s$^{-1}$.   

\section{THE SPHERICALLY SYMMETRIC WIND}

To compute the surface brightness, $I_{\nu}$, of the nebulosity, we numerically determine:
\begin{equation}
I_{\nu}\;=\;{\int}\,{\chi}_{\nu}\,{\rho}(D)\,B_{\nu}(T_{gr})\,ds
\end{equation}
where $B_{\nu}$ is the Planck function, $ds$ is the increment of distance 
along the line of sight,  ${\rho}(D)$ is the dust density at distance, $D$, from the star and the dust  opacity is ${\chi}_{\nu}$.  
To compute the dust temperature as a function of the distance, we follow Sopka et al. (1985) and assume that the dust grains are heated by the light from the central star and re-radiate in the infrared.  For carbon-rich particles, we adopt a simple model
for the grain opacity, ${\chi}_{\nu}$ (cm$^{2}$ g$^{-1}$) such that
\begin{equation}
{\chi}_{\nu}\;=\;{\chi}_{{\nu}_{0}}\;({\nu}/{\nu}_{0})^{1.3}
\end{equation}
with  ${\chi}_{{\nu}_{0}}$ = 150 cm$^{2}$ g$^{-1}$ at ${\nu}_{0}$ = 5 ${\times}$ 10$^{12}$ Hz (60 ${\mu}$m)  (Le Bertre 1997).  When the circumstellar envelope is optically thin,  the dust grain temperature, $T_{gr}$, can be written as:
\begin{equation}
T_{gr}\;=\;0.77\,T_{*}\,(R_{*}/D)^{0.38}
\end{equation}
With this prescription, $T_{gr}$
at 1${\arcsec}$ from the central star is 170 K while  Meixner et al. (1997)
and  Dayal et al. (1998) inferred values of 160 K and 185 K, respectively.
For  11.7 ${\mu}$m photons, $h{\nu}/k$ = 1230 K, and therefore  the nebulosity is produced by grains emitting on the Wien portion of the Planck curve.      

To compute ${\rho}(D)$, we assume a spherically symmetric envelope where the dust outflow velocity, $V_{gr}$, does not change with time.  We assume that the mass loss rate of grains, $\dot{M}_{gr}(t)$, and $T_{*}$ were constant until a moment came when the star's effective temperature started to increase and the wind started to die.  During the dying-wind phase, we assume that $dT_{*}/dt$ was constant.  The current boundary between
the zone where the wind had a constant mass loss rate and where it started
to die, lies at a distance $D_{die}$ with $D_{die}$ = $V_{gr}$ $t_{die}$.  Here, $t_{die}$ denotes the amount of time that has elapsed since that moment when the  star's effective temperature began to increase.  
From the equation of continuity and the simple mapping between the dust travel time, $t_{travel}$, and
distance from the star, then:
\begin{equation}
{\rho}_{gr}(D)\;=\;{\dot{M}}_{gr}(t_{travel})/(4\,{\pi}\,V_{gr}\,D^{2})
\end{equation} 
In the models for mass loss from carbon-rich AGB stars by Arndt et al. (1997), $\dot{M}_{gr}$ varies as $T_{*}^{-8.26}$.  Such a steep variation of the mass loss
with temperature occurs because the mass loss rate is controlled by radiation pressure on grains, and the dust formation is very sensitive to the gas temperature.  
   
The best  fit to the 11.7 ${\mu}$m image shown in Figure 1 is with a model derived from equation (1) where  $D_{die}$ = 5 ${\times}$ 10$^{16}$ cm, corresponding to an angular radius of 1{\farcs}5 or $t_{die}$ = 1500 yr, and $dT_{*}/dt$ = +2.2 K yr$^{-1}$.  We show in Figure 2 the results of  comparing the results from this model with transverse cuts through the intensity data at 11.7 ${\mu}$m.    In the region where $D$ $<$ $D_{die}$, the projected 11.7 ${\mu}$m intensity depends upon the multiplication of two rapidly varying quantities: the mass loss rate and the Wien portion of the Planck function.  The qualitative
agreement between our data and the observations lends support to the dying wind model, but other possible models for ${\dot{M}}(t)$ are not
excluded by our observations.   

At 1${\farcs}$5 from the star, the 11.7 ${\mu}$m intensity of the nebula is about 2.0
 Jy arcsec$^{-2}$.  Therefore, in the outer region, $\dot{M}_{gr}$ =  1.7 ${\times}$ 10$^{-7}$ M$_{\odot}$ yr$^{-1}$.     Although uncertain,  a dust to gas ratio of 5 ${\times}$ 10$^{-3}$ is plausible and therefore the total mass loss rate before the wind started to die was ${\sim}$3 ${\times}$ 10$^{-5}$ M$_{\odot}$ yr$^{-1}$. 
Given the uncertainties in the distance to the star, the dust to gas ratio, the emissivity of the grains and the observations that there is an appreciable amount of carbon in C I as well as CO (Bakker et al. 1996, Bakker \& Lambert 1998, Knapp et al. 2000), this estimate of the mass loss rate for the phase just before the winds dies is consistent with previous estimates by Hrivnak et al. (1989), Meixner et al. (1997) and Dayal et al. (1998).    
  
\section{DISCUSSION}
As a first approximation, much of our data can be explained in terms of
a nearly-spherical wind of ${\sim}$3 ${\times}$ 10$^{-5}$ M$_{\odot}$ yr$^{-1}$ which started to die about 1500 yr ago, consistent with models for post-AGB evolution of carbon-rich stars (Vassiliadis \& Wood 1993, Blocker 1995, Schroder et al. 1999).  

In order to produce a high enough pressure in the outflow to manufacture TiC, von Helden et al. (2000) proposed a short-lived phase with a steady-state mass loss rate of 10$^{-3}$ M$_{\odot}$ yr$^{-1}$  which is not supported by our data.  Nevertheless, models for shock-driven  mass loss from AGB stars may possibly explain the  presence of TiC around HD 56126.  
In  calculations by 
Winters et al. (1997) for a carbon star with a time-averaged mass loss rate of 1.2 ${\times}$ 10$^{-4}$ M$_{\odot}$ yr$^{-1}$, as shown in their Figure 2,  at 2 stellar radii above the star, the density can be as high as 10$^{-11}$ g cm$^{-3}$ when the gas temperature is 1800 K.  Therefore, the gas pressure can be  ${\sim}$1 dyne cm$^{-2}$  at a temperature nearly low enough to  produce solid TiC (Bernatowicz et al. 1996).  
There may have been phases during the pulsational cycle of HD 56126 while it was on the AGB when the physical conditions were favorable for the production of the apparent carrier
of the 21 ${\mu}$m emission feature even if the mass loss rate was as ``low" as 3 ${\times}$ 10$^{-5}$ M$_{\odot}$ yr$^{-1}$.    

The nebula shows significant deviations from spherical symmetry.  Because the mass loss
rate is extremely sensitive to the photospheric temperature, we tentatively
explore the possibility that deviations from spherical symmetry in the dust
shell can be explained by temperature variations on the surface of the star
while it was losing mass.  

\subsection{The Outer Ellipticity}
As can be seen in Figure 1, beyond 1{\farcs}5 from the center, for a given distance from the center, the intensity at position angle 45$^{\circ}$ is greater than at, say, position angle 135$^{\circ}$.   From equation (1), if the grain opacity is the same throughout the nebula, then these data can be reproduced either by variations in ${\rho}$ or by variations in $T_{gr}$.   The 18.7 ${\mu}$m image
displays  the same ellipticity as does the 11.7 ${\mu}$m image so that azimuthal variations in the temperature do not seem to explain the variations in the intensity.  Instead, we propose that there are more grains at position angle 45$^{\circ}$ than at position angle 135$^{\circ}$ and thus that
 there was more mass loss
directed toward position angle 45$^{\circ}$.

If the star is rotating and there is enhanced mass loss at the equatorial bulge (Dorfi \& Hofner 1996, Soker 1998, Reimers, Dorfi \& Hofner 2000), then qualitatively, the data can be explained if the rotation axis is projected
along position angle 135$^{\circ}$.  If ${\dot {M}}$ varies as $T_{*}^{-8.26}$ then a 1\% variation in the temperature can lead to an 8\% variation in the mass loss rate -- a change
great enough to explain the observed 7\% ellipticity in the isophotes. 
     
If  the surface of the
star is an equipotential, if the star rotates as a solid body with
angular velocity ${\omega}_{*}$, and if ${\theta}$ denotes the co-latitude, then:
\begin{equation}
(1/2)\,{\omega}_{*}^{2}\,R_{*}({\theta})^{2}\,sin^{2}\,{\theta}\;+\;
GM_{*}/R_{*}({\theta})\;=\;constant
\end{equation}
 When the star was on the AGB with   $T_{*}$ = 2600 K, ${\overline R_{*}}$ = 2.8 ${\times}$ 10$^{13}$ cm and $M$ = 0.6 M$_{\odot}$, and if $R_{eq}/R_{pole}$ was ${\sim}$1.02, consistent with a temperature contrast between the poles and equator of 1.01 and the assumption that $R^{2}T_{*}^{4}$ is constant over the surface of the star, then ${\omega}_{*}$ was approximately 1 ${\times}$ 10$^{-8}$ s$^{-1}$.  The angular momentum of the rotating AGB
star was approximately 2/9 $M_{*}\,R_{*}^{2}\,{\omega}_{*}$ (Soker 1998) or 2 ${\times}$ 10$^{51}$ g cm$^{2}$ s$^{-1}$.  It is plausible that HD 56126 could have acquired this angular momentum if, when it first swelled onto the AGB, it engulfed a putative 0.1 M$_{\odot}$ companion in a circular orbit at 1 AU which would then have donated 9 ${\times}$ 10$^{51}$ g cm$^{2}$ s$^{-1}$ of angular momentum. 
  An example of a carbon
star that has probably engulfed a companion and acquired such a large amount of angular momentum is V Hya (Barnbaum, Morris \& Kahane 1995, Kahane et al. 1996). However, we have no  evidence in support of the proposed magnitude or orientation of the rotation of HD 56126, and our hypothesis to explain the observed ellipticity is quite tentative.
  
\subsection{The Inner Blobs}
Meixner et al. (1997) propose that the bright blobs are the consequence of
an equatorially-enhanced ejection.  Their model implies that the position angle of the rotation axis lies  near 45$^{\circ}$; a very different orientation from what we propose. One difficulty with the model by Meixner et al. (1997) is that the southeastern blob is considerably brighter than the northwestern blob.  However, it is possible that the  bright blob is a factor of 1.2 closer to the
central star than the faint blob, and this could explain the intensity difference between the two blobs.    An offset of the central star from the center of the nebula could
occur if the system is a wide binary as apparently occurs in TT Cyg (Olofsson et al. 2000).  

 If, as we propose above,  the position
angle of the rotation axis is 135$^{\circ}$, then another
model besides that of Meixner et al. (1997) is required to explain the existence of the bright blobs.  
One possibility  is  that as the star first evolved away from the AGB, the magnetic field was strongest at the rotational poles with the consequence that the gas was particularly cold  in these regions.   We speculate that the mass loss  continued from the poles for 500 years after it had greatly diminished from the equator. If the field were stronger at one pole than the other, then the two blobs in the nebula could have different intensities.  To appreciably change the surface temperature of the star,   the energy density in the  magnetic field, $B^{2}/(8{\pi})$ must be comparable to the thermal pressure in the photosphere (Soker \& Clayton 1999).    The required value of the magnetic field is therefore $[(16\,{\pi}g)/(3{\chi})]^{1/2}$
where ${\chi}$ is the Kramers opacity in the atmosphere or ${\sim}$ 3 ${\times}$ 10$^{-3}$ cm$^{2}$ g $^{-1}$ (Soker 1998) and $g$ is the gravitational acceleration which was near 0.1 cm s$^{-2}$ when the star was just leaving the AGB. Thus a field of ${\sim}$20 Gauss is implied. From VLBA maps of polarized SiO masers, magnetic fields approaching this amplitude have been inferred around the oxygen-rich Mira variable TX Cam (Kemball \& Diamond 1997), but the true value of the magnetic field is controversial and may be only 0.03 Gauss (Wiebe \& Watson 1998,  Desmurs et al. 2000). 

\section{CONCLUSIONS}

We have obtained high-resolution mid-IR images of the detached dust shell around
the carbon star HD 56126 and
propose the following:

1.  The gross features of the region beyond a radius of 1${\farcs}5$ can be reproduced by an approximately spherical 
 wind from a star which had a  mass loss rate of ${\sim}$3 ${\times}$ 10$^{-5}$ M$_{\odot}$  yr$^{-1}$ while it was on the AGB.  
After the star left the AGB, its effective temperature increased with an average rate of +2.2 K yr$^{-1}$ and the mass loss
rate varied as $T_{*}^{-8.26}$, as expected in models for dust-driven winds.

2.  The extreme sensitivity of the mass loss rate to the effective temperature of the star allows us to tentatively propose that both the ellipticity in the outer envelope and the inner blobs can be explained by plausible temperature variations on the surface of the star which might have resulted
from rotation and/or large scale magnetic fields.   

We thank M. Meixner for a particularly thoughtful referee's report and M. Morris for his comments.
This work has been partly supported both at UCLA by NASA and at the Jet Propulsion Laboratory, California Institute of Technology, under an agreement with NASA.

\newpage
\begin{center}
{\bf FIGURE CAPTIONS}
\end{center}
Fig 1. The 11.7 ${\mu}$m image of HD 56126.  North is to the top and and East is to 
the left; these directions are displayed on the Figure by lines whose 
lengths are 1${\arcsec}$ each.  The maximum surface brightness that we detected and shown in the red color is 4.0 Jy arcsec$^{-2}$.  The yellow, green and purple
colors correspond to 2.4, 1.6 and 0.40 Jy arcsec$^{-2}$, respectively.
\\
\\
Fig. 2.  Two cuts through the 11.7 ${\mu}$m image of HD 56126.    Red and green crosses are used for tracing the images at position angles of 135$^{\circ}$ and 45$^{\circ}$, respectively, while the results from the spherically symmetric model are plotted as a solid blue line.
\end{document}